\documentclass[showpacs,floatfix]{revtex4-1}
\usepackage{graphicx,psfrag,amsmath,amssymb,amsfonts,latexsym,color,epsf,graphpap}

\begin{document}

\title{The effect of concurrent geometry and roughness in interacting
surfaces}
\author{C.~D.~Fosco$^{a}$, F.~C.~Lombardo$^{b}$, F.~D.~Mazzitelli$^{a}$}
\affiliation{$^a$Centro At\'omico Bariloche and Instituto Balseiro, 
Comisi\'on Nacional de Energ\'\i a At\'omica, R8402AGP Bariloche, Argentina.\\
$^b$Departamento de F\'\i sica {\it Juan Jos\'e
Giambiagi}, FCEyN UBA and IFIBA CONICET-UBA, Facultad de Ciencias Exactas y Naturales,
Ciudad Universitaria, Pabell\' on I, 1428 Buenos Aires, Argentina}

\date{today}

\begin{abstract} 
We study the interaction energy between two surfaces, one of them flat, the
other describable as the composition of a small-amplitude corrugation and a
slightly curved, smooth surface. The corrugation, represented by a
spatially random variable, involves Fourier wavelengths shorter than the
(local) curvature radii of the smooth component of the surface.  After
averaging the interaction energy over the corrugation distribution, we
obtain an expression which only depends on the smooth component. We then
approximate that functional by means of a derivative expansion, calculating
explicitly the leading and next-to-leading order terms in that
approximation scheme.  We analyze the resulting interplay between shape and
roughness corrections for some specific corrugation models in the cases of
electrostatic and Casimir interactions.  
\end{abstract}
\maketitle
\section{Introduction}\label{sec:intro}
The computation of interaction forces between close surfaces has been the
subject of interest in many areas of physics, like colloidal and
macromolecular phenomena, nuclear physics, electrostatics, van der Waals
and Casimir interactions. 

For the case of two surfaces which are `smooth', namely, that have curvature radii
much larger than the typical distance between them, there is a time-honored
analytical tool to compute the total interaction force, the so-called
Proximity Force Approximation (PFA) \cite{Derjaguin}. This tool has, as its main virtue,
that of delivering an answer in a rather straightforward
fashion, usually in the form of an ordinary integral. Besides, it only
demands two ingredients, the first being the knowledge of the force that would
result in the same system if the two surfaces were infinite parallel plates (the simplest
possible geometry). The second ingredient is the actual geometry of the two
surfaces, i.e., part of the definition of the problem itself. 

As it may be inferred from the previous remarks, the PFA satisfies a kind of
`universality', since the approach is essentially geometric. In fact, the
underlying microscopic mechanism responsible for the force is only required
at the stage of determining the force between parallel plates, the rest
being dictated solely by the geometry of the surfaces. In other words, the
PFA cannot distinguish (except for a global factor) between interactions
which give place to the same force law between parallel plates.  

Although it was originally presumed that the PFA should work reasonably well for close
and gently curved surfaces, it has generally been assumed to be an
uncontrolled approximation.  Therefrom stemmed its major drawback: the
absence of a procedure to asses its accuracy, since,
for instance, a procedure to evaluate next to leading order (NTLO)
corrections was lacking.  In an attempt to deal with that limitation, in
the last years we proposed, tested, and applied a new approximation scheme,
the so-called Derivative Expansion (DE). Although originally introduced
within the scope of the Casimir effect, for the calculation of the
interaction energy between two smooth surfaces~\cite{Foscoetal2011}, this
approximation has been shown to be a natural extension of the PFA, and it
has proven to be useful in rather different situations, not just for
Casimir effect calculations  \cite{Foscoetal2014}. The DE provides a systematic way of justifying
the PFA and, under some circumstances, also of evaluating the NTLO
corrections. One of the features that make the DE
appealing is that the corrections are also geometric in nature, as in the
PFA, and the result for the interaction can also be expressed as an
integral. On the other hand the main difference is, naturally enough, that
knowledge of the interaction for the case of parallel plates is not 
sufficient to determine the NTLO.

Although the DE approach may, and in fact has, been extended along many
different directions \cite{pfaemig,pfaemig2,pfaT,pfaem}, we shall consider here a generalization that may
appear, at first sight, to be unnatural, since it corresponds to rough (in
opposition to smooth) surfaces. However, we have in mind applying the DE to
the smooth surfaces on top of which a rough component exists.

The influence of roughness on the interaction between surfaces has been
studied in different contexts, mostly for cases where it has the form of a
ripple on top of otherwise plane surfaces. This is a problem which
 arises, for example, in the electrostatic interaction framework,
when calculating the capacitance of two (quasi) flat electrodes, to show
that it increases in the presence of roughness~\cite{Zhao}. There is also
an  observable influence of roughness \cite{Mostep}
and periodic corrugations \cite{Intravaia2013,Banishev2013} in Casimir
forces.  In situations where the amplitude of roughness
is much smaller than the distance between surfaces, the effect of roughness
can be dealt with perturbatively \cite{li91,Neto2005,Schaden2014}, taking as zero order the interaction
resulting for the smooth flat surfaces (defined to be the spatial averages of
the rough ones).  

There have been some attempts to analyze the concomitant effect of shape and
roughness corrections on the interaction between surfaces. For instance,
the particular case in which the asperities of the surface contains occasional high peaks
and  deep troughs, in addition to small-scale roughness, has been considered
in Ref.~\cite{Broer 2011}. The theoretical approach consisted in a
perturbative evaluation of the small scale roughness combined with PFA for
the large scale peaks and troughs.  A related situation has been
considered in Ref.~\cite{Bimonte2013}. The system considered there consists
in surfaces whose local separation is the sum of a slowly varying component
due to overall shape, and a rapid part due to modulation. The
calculations were performed using the PFA, and the approximations performed
are tantamount to consider a two-step PFA, in which the effect of the rapid
modulation is first considered for parallel plates, and the resulting
interaction per unit area is used in a second step to take into account the
slow changes in the overall shape. A similar geometry was considered
in Ref.~\cite{Banishev2013}, both from the theoretical and experimental points
of view. The effect of the modulation was computed using the DE, and then the slowly
varying shape using PFA. Interestingly enough, the experimental results confirm
the NTLO corrections computed using the DE.

The aim of this paper is to compute the shape and roughness corrections
systematically, looking for the existence of combined effects  between
them, beyond the PFA.
Specifically, we will consider  systems in which the geometry can be
specified by defining just two surfaces: one of them, to be denoted by $L$,
is assumed to be a plane, which (by a proper choice of orthogonal Cartesian
coordinates), corresponds to the equation $x_3 = 0$ (we use $x_1$, $x_2$
and $x_3$ to denote such a choice). In terms of the same set of coordinates
the other surface, $R$, can  be defined as follows: 
\begin{equation}\label{eq:defr}
x_3 \;=\; \Psi({\mathbf x_\shortparallel}) \;\;,\;\;\;\;
{\mathbf x_\shortparallel} \equiv (x_1,x_2) \;.
\end{equation}
An important difference between this and previously considered applications
of the DE, is that the function $\Psi({\mathbf x_\shortparallel})$ will not
be regarded as necessarily smooth; rather, we characterize it by the
property that it can be decomposed as the sum of two terms:  
\begin{equation}\label{eq:Psi}
\Psi({\mathbf x_\shortparallel}) \,=\, \psi({\mathbf x_\shortparallel}) \,+\,
\xi({\mathbf x_\shortparallel}) \;,  
\end{equation}
where $\psi({\mathbf x_\shortparallel})$ is a smooth function, while
$\xi({\mathbf x_\shortparallel})$ takes into account the `rough' component of
the surface.

We cannot apply the DE to non-smooth surfaces, but we will first average
the interaction energy over the corrugation $\xi({\mathbf
x_\shortparallel})$, and afterwards take into account the nontrivial
geometry, characterized by the smooth function $\psi({\mathbf
x_\shortparallel})$, this time by a legitimate use of the DE. Going beyond
the leading order (PFA),  we will be able to asses the relative weight of
the roughness and shape corrections, including in the theoretical
description the combined effect of both. 

This paper is organized as follows: in Section~\ref{sec:de} we derive
general expressions for the leading and NTLO terms, for the DE of the
functional representing the interaction energy between the two surfaces, up
to the first non-trivial order in the correlation function $\Omega$.
Section~\ref{sec:general} contains a discussion of some general properties
of the novel DE that contains the effect of roughness.  Then in
Section~\ref{sec:electro}, we apply our general results to the case of the
electrostatic interaction between two metallic surfaces
at fixed electrostatic potentials.  Section~\ref{sec:casimir} deals with
the DE for the interaction energy in the case of the Dirichlet and Neumann
Casimir effects.  Finally,  in Section~\ref{sec:concl} we present our conclusions.  
\section{Derivative expansion for the interaction energy}\label{sec:de}
\subsection{Averaging out the corrugation}\label{ssec:aver}
The corrugation function $\xi({\mathbf x_\shortparallel})$, introduced in
Eq.(\ref{eq:Psi}), will be regarded as a random variable with null average:
\begin{equation}\label{eq:zeroav}
\langle \xi({\mathbf x_\shortparallel}) \rangle \;=\;0 \;,
\end{equation}
where $\langle \ldots \rangle$ denotes statistical average.  Condition
(\ref{eq:zeroav}) can always be achieved, by a constant shift in $\psi$.
For $n$ spatial arguments: ${\mathbf x^{(i_1)}_\shortparallel}, \ldots,
{\mathbf x^{(i_n)}_\shortparallel}$, statistical averages are defined by
the functional integral:
\begin{equation}
\langle 
\xi({\mathbf x^{(i_1)}_\shortparallel}) \ldots   
\ldots 
\xi({\mathbf x^{(i_n)}_\shortparallel})
\rangle 
\;=\; 
\frac{\int {\mathcal D} \xi \;
\xi({\mathbf x^{(i_1)}_\shortparallel}) \ldots \xi({\mathbf
x^{(i_n)}_\shortparallel}) \,  
e^{- W[\xi]}}{\int {\mathcal D}\xi \, e^{- W[\xi]}} \;,
\end{equation}
where $W[\xi]$ denotes a (real) not necessarily quadratic, functional of
the corrugation function $\xi({\mathbf x_\shortparallel})$, determining the
statistical weight of the different configurations.

Since we assume roughness to be small, to the first non trivial order in
its amplitude, we shall only need the two-point autocorrelation function
$\Omega$:  
\begin{equation}
\Omega({\mathbf x_\shortparallel},{\mathbf y_\shortparallel})  \;\equiv\;
\langle \xi({\mathbf x_\shortparallel}) \xi({\mathbf y_\shortparallel}) \rangle \;, 
\end{equation} 
since autocorrelation functions involving more than two $\xi$'s are
of higher order in the amplitude. Note that, because of Eq.(\ref{eq:zeroav}),
$\Omega({\mathbf x_\shortparallel},{\mathbf y_\shortparallel})$ is a
connected function of its arguments.
In the quadratic $W[\xi]$ case,
\begin{equation}\label{eq:quadw}
W[\xi]\;=\; \frac{1}{2} \int_{{\mathbf x_\shortparallel},{\mathbf y_\shortparallel}}
\xi({\mathbf x_\shortparallel}) M({\mathbf x_\shortparallel},{\mathbf
y_\shortparallel})
\xi({\mathbf y_\shortparallel})\;,
\end{equation}
where $M$ is a positive-definite kernel. In this case, one has
\begin{equation}
\Omega({\mathbf x_\shortparallel},{\mathbf y_\shortparallel})\;=\; 
M^{-1}({\mathbf x_\shortparallel},{\mathbf y_\shortparallel}) \;.
\end{equation}

At this point, we introduce an extra requirement, involving both
$\Omega$ and $\psi$, the smooth component of $\Psi$: the autocorrelation
length $l$, determined by $\Omega$, must be much smaller than the curvature
radius of the smooth surface defined by $\psi$. Then, whenever the
autocorrelation function differs appreciably from zero, the smooth surface
can be regarded as approximately flat. In geometrical terms, the relevant
pairs of arguments of $\Omega$ lie on the same tangent plane to each point
of the $R$ surface. Hence, we
only need to know the function $\Omega$, as an ingredient in the
forthcoming calculations, for both arguments lying on the same plane
surface, where it is translation-invariant (we assume the surfaces
$R$ and $L$ to be homogeneous).  Thus, we approximate $\Omega({\mathbf
x_\shortparallel},{\mathbf y_\shortparallel})$ as follows:
\begin{equation}
\Omega({\mathbf x_\shortparallel},{\mathbf y_\shortparallel}) \;\simeq\;
\Omega({\mathbf x_\shortparallel}-{\mathbf y_\shortparallel}) \;=\;
\Omega(|{\mathbf x_\shortparallel}-{\mathbf y_\shortparallel}|) \;,
\end{equation}
where the last equality follows from the (assumed) local isotropy of the curved
surface.

The interaction between the two surfaces is  a functional, $F[\Psi]$.
However, we are not interested in the detailed dependence on $\Psi$, since
that function contains the random, rapidly varying component, $\xi$. Our
first step is then to average out the dependence on $\xi$ over its statistical
distribution, obtaining a functional $F_{\rm  eff}$, depending on $\psi$,
the smooth part of $\Psi$:
\begin{equation}\label{eq:deffeff}
F_{\rm  eff}[\psi] \;\equiv\; \langle F[ \psi + \xi] \rangle \;.
\end{equation}
On the other hand, since the amplitude of $\xi$ is assumed to be small, we
shall functionally expand $F$ in powers of $\xi$:
\begin{eqnarray}\label{eq:feff1}
F_{\rm  eff}[\psi] &=& \langle F[ \psi ] \,+\, \int_{{\mathbf x_\shortparallel}}
F^{(1)}_\psi({\mathbf x_\shortparallel}) \xi({\mathbf x_\shortparallel}) 
+\frac{1}{2}\, \int_{{\mathbf x_\shortparallel},{\mathbf y_\shortparallel}} 
F^{(2)}_\psi({\mathbf x_\shortparallel},{\mathbf y_\shortparallel}) \xi({\mathbf x_\shortparallel}) 
\xi({\mathbf y_\shortparallel})  \nonumber\\
&+& \frac{1}{3!}\, \int_{{\mathbf x_\shortparallel},{\mathbf
y_\shortparallel},{\mathbf z_\shortparallel}}
F^{(3)}_\psi({\mathbf x_\shortparallel},{\mathbf y_\shortparallel},{\mathbf
z_\shortparallel})\, \xi({\mathbf x_\shortparallel}) \xi({\mathbf y_\shortparallel}) 
\xi({\mathbf z_\shortparallel})\rangle
\,+\,\ldots
\nonumber\\
&=&  F_{\rm  eff,0}[ \psi ] \,+\, F_{\rm  eff,2}[\psi] \,+\,F_{\rm  eff,3}[\psi]\,+\,\ldots\;,
\end{eqnarray}
where the subindex $k$ in $F_{\rm eff,k}$ denotes $k^{th}$ order in $\xi$, the rugosity
amplitude, and we have introduced the notation:
\begin{equation}
F^{(n)}_\psi({\mathbf x^{(1)}_\shortparallel},\ldots, {\mathbf
x^{(n)}_\shortparallel}) \;\equiv\;
\frac{\delta^n F[\chi]}{\delta \chi({\mathbf x^{(1)}_\shortparallel})\ldots
\delta \chi({\mathbf x^{(n)}_\shortparallel})}\Big|_{\chi=\psi} \;.
\end{equation}

Clearly,
\begin{equation}\label{eq:f0}
F_{\rm  eff,0}[\psi] \;=\; F[\psi]\,
\end{equation}
and
\begin{equation}\label{eq:f2}
F_{\rm  eff,2}[\psi] \;=\; \frac{1}{2} \, \int_{{\mathbf
x_\shortparallel},{\mathbf
y_\shortparallel}} F^{(2)}_\psi({\mathbf x_\shortparallel},{\mathbf y_\shortparallel}) \,
\Omega({\mathbf y_\shortparallel} - {\mathbf x_\shortparallel})  \;.
\end{equation}
Regarding the third and higher orders, in general they will require the
knowledge of $n$-point correlation functions of the rugosity. An important
exception is the Gaussian (quadratic) weight case, where only even numbers of fields yield
non-vanishing contributions. 

In Eq.(\ref{eq:feff1}), we will focus on the first two terms: the first, $F_{\rm  eff,0}[
\psi ]$, is identical to the functional $F$ {\em in the absence of
corrugation}, and the second, $F_{\rm  eff,2}$, which contains the first nontrivial
correction due to the small-amplitude corrugation. That small amplitude
assumption justifies the inclusion of just the first order in $\Omega$ or,
equivalently, the leading nontrivial order in the corrugation amplitude. 
Let us now proceed to perform the DE, to the second order in derivatives,
for those two terms $F_{\rm eff,0}$ and $F_{\rm  eff,2}$. 

\subsection{Derivative expansion for $F_{\rm  eff}$}\label{ssec:f}
The averaged functional $F_{\rm  eff}$, when expanded in derivatives up to the
second order, should fall into an expression having the form~\cite{Foscoetal2014}
\begin{equation}\label{eq:de0}
F_{\rm  eff}[\psi] \;\simeq\; F_{\rm P}[\psi] \,+\, F_{\rm D}[\psi] \;, 
\end{equation}
where $F_{\rm P}[\psi]$, the term without derivatives, is just the PFA
approximation to $F_{\rm  eff}$: 
\begin{equation}\label{eq:deffp}
F_{\rm P}[\psi] \;=\; \int d^2{\mathbf x_\shortparallel} \, V(\psi({\mathbf
x_\shortparallel}))\; 
\end{equation} 
and $F_{\rm D}$ denotes the term with two derivatives:
\begin{equation}\label{eq:deffd}
F_{\rm D}[\psi] \;=\; \int d^2{\mathbf x_\shortparallel} \, Z(\psi) |\nabla \psi |^2 \;. 
\end{equation}

On the other hand, since $F_{\rm  eff}$ has been expanded in powers of the
corrugation, we shall have like expansions for $V$ and $Z$; namely,
\begin{equation}
V=V_0+V_2+\ldots \;\;,\;\;\; Z=Z_0+Z_2+\ldots \, ,
\end{equation}
and consequently
\begin{equation}
F_{\rm P}=F_{\rm P,0}+F_{\rm P,2}+\ldots\;\;,\;\;\;F_{\rm D}=F_{\rm D,0}+F_{\rm D,2}\, .
\end{equation}
Here, the first terms in the expansion are independent of the corrugation,
and therefore they can be obtained by applying the DE to the functional
$F_{\rm  eff,0} \equiv F$. Thus, the function $V_0$ can be determined, for example, from the value of
$F$ for the special case of the parallel plates geometry:
\begin{equation}\label{eq:v0a}
V_0(a) \;=\; \frac{F[a]}{\mathcal S} \;, 
\end{equation}
where ${\mathcal S}$ denotes the area of the $L$ plate (this factor cancels
out a similar one which appears in the numerator because of translation
invariance on the plane when the plates are flat and parallel).

On the other hand, $Z_0$ can be determined from the knowledge of
$F^{(2)}[\psi]$ for $\psi = a$ \cite{Foscoetal2011,Foscoetal2014}:
\begin{equation}\label{eq:za}
Z_0(a) \;=\; \frac{1}{4} \big[ \Delta_{\mathbf k_\shortparallel} {\widetilde
f}^{(2)}({\mathbf k_\shortparallel}) \big]_{{\mathbf k_\shortparallel} \to 0} 
\end{equation}
with $\Delta_{\mathbf k_\shortparallel} \equiv \frac{\partial^2}{\partial
k_1^2}+\frac{\partial^2}{\partial k_2^2}$, the Laplacian with respect to
the momenta parallel to the \mbox{$x_3 = {\rm constant}$} planes.  Here,
${\widetilde f}^{(n)}$ is obtained from ${\widetilde F}_a^{(n)}({\mathbf
k_\shortparallel}^{(1)},...,{\mathbf k_\shortparallel}^{(n)})$, the Fourier
transform of $F_a^{(n)}$, by a procedure which we detail now.  Since they
are expansion coefficients defined around the translation-invariant
configuration $x_3=a$, ${\widetilde F}_a^{(n)}$ must be proportional to the
delta function of momentum conservation corresponding to each order. Thus,
\begin{equation}
{\widetilde F}_a^{(n)}({\mathbf k_\shortparallel}^{(1)},...,{\mathbf
k_\shortparallel}^{(n)}) \;=\;(2\pi)^2 \, 
\delta({\mathbf k_\shortparallel}^{(1)}+...+{\mathbf k_\shortparallel}^{(n)}) \,
{\widetilde \gamma}^{(n)}({\mathbf k_\shortparallel}^{(1)},...,{\mathbf
k_\shortparallel}^{(n)}) \;,
\end{equation}
where ${\widetilde\gamma}^{(n)}$ is completely symmetric function of its
arguments.  However, the presence of the delta function means that it can
be completely determined by a function of just $n-1$ arguments. To make
that explicit, we introduce the kernels ${\widetilde f}^{(n)}$, as follows:
\begin{equation}
{\widetilde f}^{(n)}({\mathbf k_\shortparallel}^{(1)},...,{\mathbf
k_\shortparallel}^{(n-1)}) \;\equiv\;
{\widetilde \gamma}^{(n)}({\mathbf k_\shortparallel}^{(1)},...,{\mathbf
k_\shortparallel}^{(n-1)}, -\sum_{i=1}^{n-1}{\mathbf k_\shortparallel}^{(i)} ) \;,
\end{equation}
among which we have the particular one appearing in the calculation of $Z_0$:
\mbox{${\widetilde f}^{(2)}({\mathbf k_\shortparallel}) \equiv
{\widetilde\gamma}^{(2)}({\mathbf k_\shortparallel},-{\mathbf k_\shortparallel})$}.
 
The function $Z_0(\psi)$ is then determined by the relation
\mbox{$Z_0(\psi) = Z_0(a)|_{a \to \psi}$}.  

Let us now consider the DE for $F_{\rm  eff,2}$, the leading correction due to
corrugation: the  function $V_2$ is, again, conveniently obtained by evaluating
$F_{\rm  eff,2}$ for the case of parallel plates. Thus:
\begin{equation}
V_2(a) \;=\; \frac{F_{\rm  eff,2}[a]}{\mathcal S} \;. 
\end{equation}
Recalling Eq.(\ref{eq:f2}), we may give a more explicit
formula for $V_2$. Indeed, from the relation:
\begin{equation}
F_{\rm  eff,2}[a] \;=\; \frac{1}{2} \, \int_{{\mathbf
x_\shortparallel},{\mathbf
y_\shortparallel}} F_a^{(2)}({\mathbf x_\shortparallel},{\mathbf y_\shortparallel}) \,
\Omega({\mathbf y_\shortparallel} - {\mathbf x_\shortparallel})  \;,
\end{equation}
and noting that
\begin{equation}
F_a^{(2)}({\mathbf x_\shortparallel},{\mathbf y_\shortparallel})\,=\, \int
\frac{d^2{\mathbf p_\shortparallel}}{(2\pi)^2} \, e^{i {\mathbf p_\shortparallel}\cdot ({\mathbf
x_\shortparallel}-{\mathbf y_\shortparallel})}  {\widetilde
f}^{(2)}({\mathbf p_\shortparallel}) \;,
\end{equation}
we get:
\begin{equation}\label{eq:vomegares}
V_2(a) \;=\; \frac{1}{2} \, \int
\frac{d^2{\mathbf p_\shortparallel}}{(2\pi)^2} \, {\widetilde f}^{(2)}({\mathbf p_\shortparallel})
\, {\widetilde \Omega}({\mathbf p_\shortparallel})\;,
\end{equation}
where ${\widetilde \Omega}$ is the Fourier transform of the autocorrelation
function $\Omega$.

Finally, we obtain an explicit expression for $Z_2$. A rather
straightforward approach would be to apply the analogue of
Eq.(\ref{eq:za}), which was used for the functional $F$, now to the functional
$F_{\rm  eff,2}$, as defined in Eq.(\ref{eq:f2}).
Thus we see that:
\begin{equation}\label{eq:zoa}
Z_2(a) \;=\; \frac{1}{4} \big[ \Delta_{\mathbf k_\shortparallel}
{\widetilde f}_2^{(2)}({\mathbf k_\shortparallel}) \big]_{{\mathbf
k_\shortparallel} \to 0} \;,
\end{equation}
where now ${\widetilde f}_2^{(2)}$ is evaluated from the
functional expansion of $F_{\rm  eff,2}$ in powers of $\eta$, to the second order
in $\eta$. However, since $F_{\rm  eff,2}$ is defined in terms of the second
functional derivative of $F$,  it is evident that ${\widetilde
f}_2^{(2)}$  will involve the {\em fourth order\/} derivatives of $F$.
Indeed, we obtain
\begin{equation}\label{eq:zoa1}
Z_2(a) \;=\; \frac{1}{8}\, \int \frac{d^2{\mathbf p_\shortparallel}}{(2\pi)^2} \,
\big[ \Delta_{\mathbf k_\shortparallel} 
{\widetilde f}^{(4)}({\mathbf p_\shortparallel},- {\mathbf
p_\shortparallel},{\mathbf k_\shortparallel})\big]_{{\mathbf
k_\shortparallel} \to 0} \, {\widetilde \Omega}({\mathbf p_\shortparallel})\;,
\end{equation}
with
\begin{equation}
{\widetilde f}^{(4)}({\mathbf k_\shortparallel}, {\mathbf p_\shortparallel},
{\mathbf q_\shortparallel})\;=\; 
{\widetilde\gamma}^{(4)}( {\mathbf k_\shortparallel}, {\mathbf p_\shortparallel}, {\mathbf
q_\shortparallel}, - {\mathbf k_\shortparallel}-{\mathbf p_\shortparallel}-{\mathbf
q_\shortparallel}) \;.
\end{equation}

Thus, we have shown that the coefficients of the DE to second order for the
leading correction in an expansion in powers of the corrugations requires a
fourth order functional derivative of $F$. In other words, to evaluate the
second order correction to the DE,  to the leading order in the
corrugation, we need the result for the fourth order term in the expansion
of $F[a+\eta]$ in powers of $\eta$. 

Before presenting the  results that follow from the application of the
general formulas to particular cases, we shall consider some of their
general properties,  based on dimensional analysis, combined with
assumptions about the autocorrelation function $\Omega$ and about the
curved surface.
\section{General properties}\label{sec:general}
We note that, for a given interaction $F$, it will not be possible in
general to find exact analytical expressions for $F_P$ and $F_D$, except
for particular situations, or under under some simplifying assumptions.
Since the interaction is determined, the two objects to play
with are the autocorrelation function and the geometry. 

On the other hand, having in mind cases where the geometry ($\psi$) is also
given, it would be interesting to have a general formula for the DE,
depending on an arbitrary $\Omega$, but where the geometry has been `integrated out',
namely, it appears only by means of the parameters characterizing the
function $\psi$.  For the term without derivatives, this can be achieved by
using the so-called `height distribution function'. This representation can
be obtained, for example, by introducing a `$1$' inside the integral over
${\mathbf x_\shortparallel}$ in Eq.(\ref{eq:deffp}), written as follows: $1 =
\int_0^\infty da \, \delta(a - \psi({\mathbf x_\parallel}))$ (we assume
$\psi >0$ everywhere).
Thus,
\begin{equation}
F_{\rm P}[\psi] \;=\; \int_0^\infty da \, V(a) \, \sigma_{\rm P}(a) 
\label{FPsigma}
\end{equation}
where 
\begin{equation}
\sigma_{\rm P}(a) \;=\; 
\int d^2{\mathbf x_\shortparallel} \,
\delta(a - \psi({\mathbf x_\parallel})) \;. 
\end{equation}
We see that the geometry, for this term, is encoded in $\sigma_P$, while the part
dependent on the interaction and the corrugation is contained in $V(a)$.
More explicitly,
\begin{equation}
F_{\rm P}[\psi] \;=\; \frac{1}{2}\, \int \frac{d^2{\mathbf
p_\shortparallel}}{(2\pi)^2} \, {\widetilde \Omega}({\mathbf
p_\shortparallel})\, \int_0^\infty da \, {\widetilde f}^{(2)}({\mathbf
p_\shortparallel}) \, \sigma_{\rm P}(a) \;.
\end{equation}

For the term with two derivatives,  a similar procedure to the one applied
above yields:
\begin{equation}
F_{\rm D}[\psi] \;=\; \int_0^\infty da \, Z(a) \, \sigma_{\rm D}(a) 
\label{FDsigma}
\end{equation}
with
\begin{equation}
\sigma_{\rm D}(a) \;=\; 
\int d^2{\mathbf x_\shortparallel} \,
\delta(a - \psi({\mathbf x_\parallel})) |\nabla \psi({\mathbf
x_\shortparallel})|^2 \;, 
\end{equation}
where now the geometry of the curved plate appears in the $\sigma_D$
function, similar to the height distribution function, but weighted with
the square of the gradient of $\psi$. 
Thus:
\begin{equation}
F_{\rm D}[\psi] \;=\; \frac{1}{8}\, \int \frac{d^2{\mathbf p_\shortparallel}}{(2\pi)^2} \, 
{\widetilde \Omega}({\mathbf p_\shortparallel})\, 
\int_0^\infty da \, 
\big[ \Delta_{\mathbf k_\shortparallel} 
{\widetilde f}^{(4)}({\mathbf p_\shortparallel},- {\mathbf
p_\shortparallel},{\mathbf k_\shortparallel})\big]_{{\mathbf
k_\shortparallel} \to 0} \, \sigma_{\rm D}(a) \;.
\end{equation}

To simplify the forthcoming discussion, we use dimensional analysis
whenever possible. To that end, we note that the mass dimensions of the
objects appearing in the general formulas can be determined as soon as we
assume that $F$ is an energy; thus we have \mbox{$[F] = [M]$},
\mbox{${\widetilde \Omega} = [M]^{-4}$}, \mbox{$[{\widetilde f}^{(2)}]=
[M]^5$}, and \mbox{$[{\widetilde f}^{(4)}]= [M]^7$}.  Therefore, the
momentum-space Laplacian of \mbox{$[{\widetilde f}^{(4)}]$} has dimensions
$[M]^5$.  
We then rewrite $V$ and $Z$ using dimensionless
objects. For the PFA contribution, there is not much to be said for
$V_0(a)$ beyond expression (\ref{eq:v0a}); thus,
\begin{equation}
V_0(a) \;=\; a^{-3}  \; \left(\frac{a^3 F[a]}{\mathcal S}\right) 
\end{equation}
while for $V_2$, which is given by a momentum integral, we introduce a
dimensionless momentum ${\mathbf l_\shortparallel}$, attained
by the rescaling ${\mathbf p_\shortparallel} ={\mathbf l_\shortparallel}/a$:
\begin{equation}\label{eq:vomegaresd}
V_2(a) \;=\; a^{-3}  \;\; \frac{1}{2} \,\int \frac{d^2{\mathbf
l_\shortparallel}}{(2\pi)^2} \, {\widetilde g}^{(2)}({\mathbf l_\shortparallel})
\, {\widetilde \omega}({\mathbf l_\shortparallel})\;,
\end{equation}
where we have introduced the dimensionless functions:
\begin{equation}
{\widetilde g}^{(2)}({\mathbf l_\shortparallel}) \,=\,  a^5 \, {\widetilde
f}^{(2)}({\mathbf l_\shortparallel}/a) \;\;,\;\;\; {\widetilde
\omega}({\mathbf l_\shortparallel})  \,=\, a^{-4} \, {\widetilde
\Omega}({\mathbf l_\shortparallel}/a) \;.  
\end{equation}
Thus, to the order we are considering here, we have,
\begin{equation}
V(a) \;=\; a^{-3}  \;\;\left[\frac{a^3 F[a]}{\mathcal S} 
+  \frac{1}{2} \,\int \frac{d^2{\mathbf
l_\shortparallel}}{(2\pi)^2} \, {\widetilde g}^{(2)}({\mathbf l_\shortparallel})
\, {\widetilde \omega}({\mathbf l_\shortparallel}) \right]\;.
\end{equation}

Regarding $Z_2$, which involves ${\widetilde \Omega}$ and ${\widetilde
f}^{(4)}$, a similar rescaling yields:
\begin{equation}\label{eq:zomegaresd}
Z_2(a) \;=\; a^{-3}  \;\; \frac{1}{8} \,\int \frac{d^2{\mathbf
l_\shortparallel}}{(2\pi)^2} \, {\widetilde h}^{(2)}({\mathbf
l_\shortparallel})
\, {\widetilde \omega}({\mathbf l_\shortparallel})\;,
\end{equation}
where
\begin{equation}
{\widetilde h}^{(2)}({\mathbf l_\shortparallel}) \;=\;  a^5 \, \big[
\Delta_{{\mathbf k_\shortparallel}/a} 
{\widetilde f}^{(4)}({\mathbf l_\shortparallel}/a,- {\mathbf
l_\shortparallel}/a,{\mathbf k_\shortparallel}/a)\big]_{{\mathbf
k_\shortparallel} \to 0} \;.
\end{equation}

Let us consider some exact properties and results that can be obtained
about the contributions due to $V_2$ and $Z_2$, as well as about the
interplay between roughness and
geometry under some simplifying assumptions about the autocorrelation
function. Although the resulting models  will not necessarily correspond to
realistic situations, they have the advantage that some insight about the
interplay between the different causes may be elucidated more clearly. 
\begin{enumerate}
\item \underline{Sharp cutoff model}:
This model corresponds to the autocorrelation function:
\begin{equation}\label{correlcut}
{\widetilde\Omega}({\mathbf p_\shortparallel})\;=\; \frac{4\pi
\epsilon^2}{p_{\rm max}^2 - p_{\rm min}^2} \, \theta(|{\mathbf
p_\shortparallel}|-p_{\rm min}) \, \theta(p_{\rm max} - |{\mathbf
p_\shortparallel}|) \;,
\end{equation}
where $p_{\rm max} \geq  p_{\rm min} \geq 0$, and $\epsilon$ is a constant
with the dimensions of a length, which is a measure of the amplitude of the
corrugation (its rms value).  The constants $p_{\rm max}$ and $p_{\rm min}$
play the role of UV and IR cutoffs, respectively. Equivalently, they
determine the minimum and maximum correlation distances.

In this model, we see that the second order contribution to the PFA
approximation becomes:
\begin{equation}
F_{\rm P,2}=\int d^2{\mathbf x_\shortparallel} V_2 =\frac{\epsilon^2}{p_{\rm max}^2 - p_{\rm min}^2}\,  
\int_0^\infty da \, \frac{\sigma_{\rm P}(a)}{a^7} \,
\int_{a p_{\rm min}}^{a p_{\rm max}} dx \, x  {\widetilde g}^{(2)}(x) \;.\label{F2}
\end{equation}
On the other hand, for the term with two derivatives, determined by $Z_2$,
the equivalent expression is:
\begin{equation}
F_{\rm D,2}=\int d^2{\mathbf x_\shortparallel} Z_2 |\nabla \psi |^2  
=\frac{\epsilon^2}{4 (p_{\rm max}^2 - p_{\rm min}^2)}\,  
\int_0^\infty da \, \frac{\sigma_{\rm D}(a)}{a^7} \,
\int_{a p_{\rm min}}^{a p_{\rm max}} dx \, x \, {\widetilde h}^{(2)}(x) \;.
\end{equation}

\item \underline{$\delta$-like ${\widetilde \Omega}$}:
This case corresponds to a limit of the sharp-cutoff model, such that the
two cutoffs collapse to a common value $q$. Thus, the momentum-space
autocorrelation function has the following form:
\begin{equation}
{\widetilde \Omega}({\mathbf p_\shortparallel})\;=\; 2 \pi \epsilon^2 \,
\frac{\delta( |{\mathbf p_\shortparallel}|-q)}{|{\mathbf
p_\shortparallel}|} \;,
\end{equation}
again, $\epsilon$ is a length, which may be interpreted as a measure of the
amplitude of the corrugation, while ${\mathbf q_\shortparallel}$ determines
its momentum scale. Thus, 
\begin{equation}\label{eq:v2deltaq}
\int d^2{\mathbf x_\shortparallel} V_2 \;=\; 
\frac{\epsilon^2}{2}\,  
\int_0^\infty da \, \sigma_{\rm P}(a) \,{\widetilde f}^{(2)}(q) 
\,=\,
\frac{\epsilon^2}{2}\,  
\int_0^\infty da \, \frac{\sigma_{\rm P}(a)}{a^5} \,{\widetilde g}^{(2)}(q a)\;
\end{equation}
and
\begin{equation}
\int d^2{\mathbf x_\shortparallel} Z_2 |\nabla \psi |^2 \;=\; 
\frac{\epsilon^2}{8}\,  
\int_0^\infty da \, \sigma_{\rm D}(a) \,{\widetilde g}^{(2)}(q) 
\,=\,
\frac{\epsilon^2}{8}\,  
\int_0^\infty da \, \frac{\sigma_{\rm D}(a)}{a^5} \,{\widetilde h}^{(2)}(q a)\;.
\end{equation}

\item \underline{Momentum-independent correlation function}:

Finally, this situation corresponds to a case where the IR cutoff
tends to zero and the UV one to infinity.  Equivalently, there is a
vanishing correlation length. Thus
\begin{equation}
{\widetilde \Omega}({\mathbf k_\shortparallel})\;=\; 
{\widetilde \Omega}_0 \,\equiv\,{\rm constant} \;,
\end{equation}
where ${\widetilde \Omega}_0$ has the dimensions of  $({\rm length})^4$.
Note that this value of ${\widetilde \Omega}_0$ may be thought of as a
particular limit of the two cutoff case, such that the product of
$\epsilon$ by $p_{\rm max} - p_{\rm min}$ remains constant.

Then we see that:
\begin{equation}\label{eq:v2constomega}
V_2(a) \;=\; \frac{{\widetilde\Omega}_0}{2 a^7} \;
\int \frac{d^2{\mathbf l_\shortparallel}}{(2\pi)^2} \, {\widetilde g}^{(2)}({\mathbf
l_\shortparallel})\,.
\end{equation}
and
\begin{equation}\label{eq:z2constantomega}
Z_2(a) \;=\; \frac{{\widetilde\Omega}_0}{8 a^7} \;\int \frac{d^2{\mathbf
l_\shortparallel}}{(2\pi)^2} \, 
{\widetilde h}^{(2)}({\mathbf l_\shortparallel})\;.
\end{equation}
For the particular case of interactions which do not introduce any
dimensionful quantity into the problem (except $a$), like in the Casimir
effect for a scalar field with Dirichlet or Neumann conditions, or even the
case of the electromagnetic field, the dimensionless kernels which appear
integrated above are just numbers. Thus, in those cases we shall have,
regarding the dependence with a,
\begin{equation}\label{eq:dependa}
V_2(a) \;\propto\;  \frac{1}{a^7}\;\;,\;\;\; Z_2(a) \;\propto\;
\frac{1}{a^7} \;.
\end{equation}
An important remark is that no boundary condition is
perfect for all momenta, therefore one should also expect an UV cutoff
to exist in ${\widetilde g}^2$ and ${\widetilde h}^2$. Thus, the momentum integrals
above shall have cutoffs (in principle, unrelated to the ones of the
autocorrelation function). This should be kept in mind, in particular, when
the integrals over ${\mathbf l_\shortparallel}$ are UV divergent. In this
situation, the behaviour in Eq.(\ref{eq:dependa}) should be reliable when $a$ is
much larger than the inverse of the UV cutoff. For small values of $a$,
however, one should expect a smoother behaviour than that of
Eq.(\ref{eq:dependa}). 
\end{enumerate} 

Finally, in order to gain some insight, and because it is representative of many
relevant situations, we evaluate the geometrical factors $\sigma_{\rm P}$ and
$\sigma_{\rm D}$ explicitly for the case of two particular surfaces which produce
relatively simple results.  

The first case we consider has revolution symmetry around the $x_3$
axis and is defined by the function $\psi = d +  b |{\mathbf
x_\shortparallel}|^\kappa$, where $d$, $b$ and $\kappa$ are positive
constants. We obtain:
\begin{equation}
\sigma_{\rm P}(a) \;=\; \left\{ \begin{array}{ccc}
0 & {\rm if} & a < d \\
\frac{2\pi}{b \kappa}  (\frac{a-d}{b})^{\frac{2}{\kappa}-1} &{\rm if}& a \geq d 
\end{array}
\right.
\end{equation}
and
\begin{equation}
\sigma_{\rm D}(a) \;=\; \left\{ \begin{array}{ccc}
0 & {\rm if} & a < d \\
2\pi \kappa  (a-d) & {\rm if} & a \geq d 
\end{array}
\right. \;.
\end{equation}

The other case corresponds to a sphere of radius to be $R$ and its distance
of closest approach to the flat surface to be $d$, we find:
\begin{equation}
\sigma_{\rm P}(a) \;=\; \left\{ \begin{array}{cc}
2 \pi R (1 - \frac{a}{R} + \frac{d}{R}) &{\rm if} \; d \leq a \leq R+d \\
0 & {\rm otherwise}
\end{array}
\right.
\end{equation}
and
\begin{equation}
\sigma_{\rm D}(a) \;=\; \left\{ \begin{array}{cc}
2 \pi (a-d) (1 + \frac{1}{1-\frac{a}{R}+\frac{d}{R}}) &{\rm if} \; d \leq a
\leq R+d -\sqrt{R^2-\rho_M^2}\\
0 & {\rm otherwise}
\end{array}
\right.
\end{equation}
where $\rho_M <R$ is a spatial cutoff, required for the term with two
derivatives to discard contributions where the approximation is not valid
(since the derivative of $\psi$ diverges). 

In the next sections, we apply the previous general expressions for the DE
corresponding to $F_{\rm  eff,2}$ to two interesting cases: the electrostatic
interaction of two metallic surfaces held at fixed electrostatic potentials
(Section~\ref{sec:electro}), and  the Casimir interaction between two
Dirichlet or Neumann surfaces acting on a quantum massless real scalar field
(Section~\ref{sec:casimir}).  
\section{Results for the electrostatic case}\label{sec:electro}

In this section, we apply the general results for the interaction energy between surfaces to the particular case of the electrostatic interaction between 
two perfect conductors
held  at a potential difference $V$. The electrostatic energy will be denoted by $U$ and is given by
\begin{equation}
U[\Psi]= \frac{\epsilon_0V^2}{2}\int d^2{\mathbf
x_\shortparallel}\int_0^{\Psi}dx_3\vert\nabla\phi\vert^2\, ,
\label{defelect}
\end{equation}
where $\phi({\mathbf x_\shortparallel},x_3)$ is the electrostatic potential, that satisfies the Laplace equation between plates, subjected to the 
boundary conditions $\phi({\mathbf x_\shortparallel},0)=0$ and $\phi({\mathbf x_\shortparallel},\Psi)=V$. As described in previous sections, in order to
obtain the derivative expansion for $U_{eff}[\psi]=<U[\psi+\xi]>$ it is necessary to compute the fourth order functional derivative of $U$ at $\Psi=a$. 
Therefore we write
\begin{equation}
\Psi=a+\eta({\mathbf x_\shortparallel})
\end{equation}
and expand the electrostatic energy up to the fourth order in $\eta$. In a previous work \cite{annphys}, we performed this calculation up to second order. We will follow
a similar approach here, extending the results to the fourth order case. 

We expand the boundary condition on the curved surface in powers of $\eta$:
\begin{equation}
V=\phi({\mathbf x_\shortparallel},a) + \eta ({\mathbf x_\shortparallel})\partial_3 \phi({\mathbf x_\shortparallel},a)+\frac{\eta^2 ({\mathbf x_\shortparallel})}{2}\partial_3^2 
\phi({\mathbf x_\shortparallel},a)+\frac{\eta^3 ({\mathbf x_\shortparallel})}{6}\partial_3^3 \phi({\mathbf x_\shortparallel},a)+....
\label{elecbc}
\end{equation}
and look for solutions of the form
\begin{equation}
\phi({\mathbf x_\shortparallel},x_3)=\sum_{n\geq 0}\phi^{(n)}({\mathbf x_\shortparallel},x_3)\, ,
\label{elecexp}
\end{equation}
where $\phi^{(n)}({\mathbf x_\shortparallel},x_3)$ is $O(\eta^n)$ and satisfies the Laplace equation. The boundary conditions are 
$\phi^{(n)}({\mathbf x_\shortparallel},0)=0$ and 
\begin{eqnarray}
\phi^{(0)}({\mathbf x_\shortparallel},a)&=&V\nonumber\\
\phi^{(1)}({\mathbf x_\shortparallel},a)&=&-\eta ({\mathbf x_\shortparallel})\partial_3 \phi^{(0)}({\mathbf x_\shortparallel},a)\nonumber\\
\phi^{(2)}({\mathbf x_\shortparallel},a)&=&-\eta ({\mathbf x_\shortparallel})\partial_3 \phi^{(1)}({\mathbf x_\shortparallel},a)
-\frac{\eta^2 ({\mathbf x_\shortparallel})}{2}\partial_3^2 
\phi^{(0)}({\mathbf x_\shortparallel},a)\nonumber\\\
\phi^{(3)}({\mathbf x_\shortparallel},a)&=&-\eta ({\mathbf x_\shortparallel})\partial_3 \phi^{(2)}({\mathbf x_\shortparallel},a)-\frac{\eta^2 ({\mathbf x_\shortparallel})}{2}\partial_3^2 
\phi^{(1)}({\mathbf x_\shortparallel},a)\, ,
\end{eqnarray}
as can be readily checked by inserting the expansion Eq.(\ref{elecexp}) into  Eq.(\ref{elecbc}).

The leading order solution is of course $\phi^{(0)}= V x_3/a$.  Taking into account the boundary condition at $x_3=0$ we have, for $n\geq 1$,
\begin{equation}
\phi^{(n)}({\mathbf x_\shortparallel},x_3)=\int \frac{d^2{\mathbf k_\shortparallel}}{(2\pi)^2} \, e^{i{\mathbf k_\shortparallel\cdot x_\shortparallel} }A^{(n)}({\mathbf k_\shortparallel})
\sinh (k_\shortparallel x_3)\, ,
\end{equation}
where the functions $A^{(n)}({\mathbf k_\shortparallel})$ are determined by the boundary conditions at $x_3=a$:

\begin{eqnarray}
A^{(1)}({\mathbf k_\shortparallel})&=&  -\frac{V}{a}\frac{\tilde\eta({\mathbf k_\shortparallel})}{\sinh (k_\shortparallel a)} \nonumber\\
A^{(2)}({\mathbf k_\shortparallel})&=&    \frac{V}{a\sinh (k_\shortparallel a)}  \int \frac{d^2{\mathbf p_\shortparallel}}{(2\pi)2} p_\shortparallel  \coth(p_\shortparallel a)
\tilde\eta({\mathbf k_\shortparallel} +{ \mathbf p_\shortparallel})\tilde\eta(-{\mathbf p_\shortparallel})   \nonumber\\
A^{(3)}({\mathbf k_\shortparallel})&=& \frac{V}{a\sinh (k_\shortparallel a)}  \int \frac{d^2{\mathbf p_\shortparallel}}{(2\pi)^2}   \frac{d^2{\mathbf p_\shortparallel}}{(2\pi)^2} 
\tilde\eta({\mathbf k_\shortparallel} +{ \mathbf q_\shortparallel})\tilde\eta({\mathbf p_\shortparallel} -{ \mathbf q_\shortparallel}) \tilde\eta(-{\mathbf p_\shortparallel})
 \nonumber\\
&&\quad\quad\quad\quad \times (p_\shortparallel q_\shortparallel \coth(p_\shortparallel a)\coth(q_\shortparallel a)-\frac{1}{2}\vert {\mathbf q_\shortparallel} -  {\mathbf p_\shortparallel}\vert^2)\, .
\end{eqnarray}

After a long but straightforward calculation, the expansion of the electrostatic energy reads
\begin{equation}
U[a+\eta]= \sum_{n\geq 0}U^{(n)}[a+\eta]
\end{equation}
with
\begin{eqnarray}
U^{(0)}&=&\frac{\epsilon_0V^2}{2a}\int d^2{\mathbf x_\shortparallel}\nonumber\\
U^{(1)}&=&-\frac{\epsilon_0V^2}{2a^2}\int d^2{\mathbf x_\shortparallel}\eta\nonumber\\
U^{(2)}&=&\frac{\epsilon_0V^2}{2 a^2}\int \frac{d^2{\mathbf k_\shortparallel}}{(2\pi)^2} k_\shortparallel\coth (k_\shortparallel a)\tilde\eta({\mathbf k_\shortparallel})
\tilde\eta(-{\mathbf k_\shortparallel})\nonumber\\
U^{(3)}&=&\frac{1}{3!}\int \frac{d^2{\mathbf k_\shortparallel}}{(2\pi)^2}\frac{ {d^2{\mathbf p_\shortparallel}}}{{(2\pi)^2}} \tilde f^{(3)}({\mathbf k_\shortparallel},{\mathbf p_\shortparallel}) \, \tilde\eta({\mathbf k_\shortparallel})\tilde\eta({\mathbf p_\shortparallel})
\tilde\eta(-{\mathbf k_\shortparallel}-{\mathbf p_\shortparallel} )\nonumber\\
U^{(4)}&=&\frac{1}{4!}\int \frac{d^2{\mathbf k_\shortparallel}}{(2\pi)^2}\frac{ {d^2{\mathbf p_\shortparallel}}}{{(2\pi)^2}} \frac{d^2{\mathbf q_\shortparallel}}{(2\pi)^2}
\tilde f^{(4)}({\mathbf k_\shortparallel},{\mathbf p_\shortparallel},{\mathbf q_\shortparallel}) \, \tilde\eta({\mathbf k_\shortparallel})\tilde\eta({\mathbf p_\shortparallel})
\tilde\eta({\mathbf q_\shortparallel})
\tilde\eta(-{\mathbf k_\shortparallel}-{\mathbf p_\shortparallel} - {\mathbf q_\shortparallel} )\, .
\label{Un}
\end{eqnarray}
The explicit expressions for the functions $\tilde f^{(n)}$ are 
\begin{eqnarray}
\tilde f^{(3)}({\mathbf k_\shortparallel},{\mathbf p_\shortparallel}) &=& - \frac{3\epsilon_0V^2}{a^2}\left({\mathbf k_\shortparallel}\cdot{\mathbf p_\shortparallel} + k_\shortparallel p_\shortparallel \coth (k_\shortparallel a)  \coth (p_\shortparallel a) \right) \nonumber\\
\tilde f^{(4)}({\mathbf k_\shortparallel},{\mathbf p_\shortparallel},{\mathbf q_\shortparallel}) &=& \frac{2\epsilon_0 V^2}{a^2}\bigg(\big (h({\mathbf k_\shortparallel},{\mathbf p_\shortparallel},{\mathbf q_\shortparallel}) + h({\mathbf k_\shortparallel},{\mathbf q_\shortparallel},{\mathbf p_\shortparallel}) + h({\mathbf p_\shortparallel},{\mathbf k_\shortparallel},{\mathbf q_\shortparallel}) \nonumber\\
&+& h({\mathbf p_\shortparallel},{\mathbf q_\shortparallel},{\mathbf k_\shortparallel}) +h({\mathbf q_\shortparallel},{\mathbf p_\shortparallel},{\mathbf k_\shortparallel}) +h({\mathbf q_\shortparallel},{\mathbf k_\shortparallel},{\mathbf p_\shortparallel})\big)\bigg)\, ,
\end{eqnarray}
with
\begin{equation}
h({\mathbf k_\shortparallel},{\mathbf p_\shortparallel},{\mathbf q_\shortparallel}) =  q_\shortparallel \coth (q_\shortparallel a)(2 q_\shortparallel^2-{\mathbf k_\shortparallel}\cdot{\mathbf q_\shortparallel} )+
p_\shortparallel q_\shortparallel\vert 
{\mathbf k_\shortparallel}+{\mathbf p_\shortparallel}\vert
\coth (p_\shortparallel a) \coth (q_\shortparallel a)\coth ( \vert {\mathbf k_\shortparallel}+{\mathbf p_\shortparallel}\vert a)\, .
\end{equation}

In order to obtain these results it is necessary to expand the electrostatic energy in powers of $\eta$, taking into account not only the expansion of the potential but also
the upper limit of integration in Eq.(\ref{defelect}).

As a partial check of our results, we can evaluate the electrostatic energy for a constant perturbation 
\begin{equation} 
\eta({\mathbf x_\shortparallel})=\eta_0\Rightarrow \tilde\eta({\mathbf k_\shortparallel})= \eta_0 (2\pi)^2\delta({\mathbf k_\shortparallel})
\end{equation}
to obtain
\begin{equation}
U\approx \frac{\epsilon_0 V^2}{2}\frac{\mathcal S}{a}\big (1-\frac{\eta_0}{a} +\frac{\eta_0^2}{a^2}-\frac{\eta_0^3}{a^3}+\frac{\eta_0^4}{a^4}\big)\, ,
\end{equation}
which is the expansion of the exact result $\epsilon_0 V^2{\mathcal S/(2 (a+\eta_0))}$ up to fourth order in the perturbation.

From the expression for $U^{(2)}$ given in Eq.(\ref{Un}) one can read the explicit form for $\tilde f^{(2)}({\mathbf k_\shortparallel})$. 
Using Eq.(\ref {eq:vomegares}) we obtain
\begin{equation}
V_2(a)=\frac{\epsilon_0V^2}{2\pi a^5}\int_0^\infty dl_\shortparallel \, l_\shortparallel^2\,  \tilde\Omega(l_\parallel/a) \coth l_\shortparallel\, \, .
\label{v2res}
\end{equation}

The function $Z_2$ can be obtained combining
the expression for $U^{(4)}$ in Eq.(\ref{Un})
with  Eq.(\ref{eq:zoa1}).  The result is
\begin{equation}
Z_2(a)=\frac{2\epsilon_0 V^2}{\pi a^5}\int dl_\shortparallel  \, l_\shortparallel \tilde\Omega(l_\shortparallel/a)B(l_\shortparallel)
\label{z2res}
\end{equation}
where
\begin{equation}
B(l_\shortparallel) = 1+(l_\shortparallel\coth l_\shortparallel+l_\shortparallel^2\coth^2 l_\shortparallel)(\frac{1}{6}+\frac{1}{8 l_\shortparallel}(\coth l_\shortparallel 
+\frac{l_\shortparallel}{\sinh ^2l_\shortparallel}(2 l_\shortparallel \coth l_\shortparallel-3)) \, .
\end{equation}

\subsection{Sphere-plane geometry}

Let us now consider the sphere-plane geometry, assuming a sharp-cutoff model for the roughness. It is easy to obtain analytic results for sufficiently large values of   $p_{\rm max} d$.  Indeed, inserting Eq.(\ref{correlcut}) into Eq.(\ref{v2res}) we obtain
\begin{equation}
V_2(a)\simeq \frac{2}{3 a^2} \epsilon_0V^2\epsilon^2p_{\rm max}\, ,
\label{v2cut}
\end{equation}
where we have assumed that the integral in Eq.(\ref{v2res}) is dominated by large values of the momentum.  A similar analysis can be done for $Z_2(a)$. Inserting Eq.(\ref{correlcut}) into Eq.(\ref{z2res}) we get
\begin{equation}
Z_2(a)\simeq \frac{1}{3a} \epsilon_0V^2\epsilon^2p_{\rm max}^2\, .
\label{z2cut}
\end{equation}

The roughness correction to the interaction energy in this geometry can be computed using
the height distribution functions introduced in the previous section
\begin{eqnarray}
U_{P,2}&=& \int_0^\infty \sigma_P(a)V_2(a)\nonumber\\
U_{D,2}&=& \int_0^\infty \sigma_D(a)Z_2(a)\, .
\end{eqnarray}
Taking the derivative of the interaction energy with respect to the sphere-plane distance $d$
one can obtain the corresponding corrections to the force, that will be denoted by ${\cal F}_{P,2}$
and ${\cal F}_{D,2}$ respectively.  The results are
\begin{eqnarray}
{\mathcal F}_{P,2}&=&-\frac{4}{3 d^2}\pi R\epsilon_0 V^2\epsilon^2p_{\rm max}\left(1-\frac{d}{R}\right)\nonumber\\
{\mathcal F}_{D,2}&=&\frac{4}{3}\pi \epsilon_0 V^2\epsilon^2p_{\rm max}^2\log(d/R)\, ,
\end{eqnarray}
where we have made an expansion for small values of $d/R$.

On the other hand, the corresponding result for smooth sphere-plane surfaces is given by
\cite{annphys}
\begin{equation}
{\cal F}_{P,0}+{\cal F}_{D,0}=-\frac{\pi\epsilon_0 V^2 R}{d}-\frac{\pi}{3}\epsilon_0 V^2 \log(d/R)\, .
\end{equation}
The ratio between the leading orders $\chi_P={\mathcal F}_{P,2}/{\mathcal F}_{P,0}$ is proportional to $\epsilon^2 p_{\rm max}/d$, while the ratio 
of the second order results $\chi_D={\mathcal F}_{D,2}/{\mathcal F}_{D,0}$ does not depend on the distance, and is proportional
to  $\epsilon^2 p_{\rm max}^2$. Note that in the large $p_{\max}$ limit the effect of roughness grows  
with the UV cutoff.  This somewhat surprising  result is also valid for periodic modulations (the corrections to PFA are larger for smaller periods), and 
can be interpreted as
due to an increase of the effective area of interaction.

\begin{figure}[!ht]
\includegraphics[width=10.0cm]{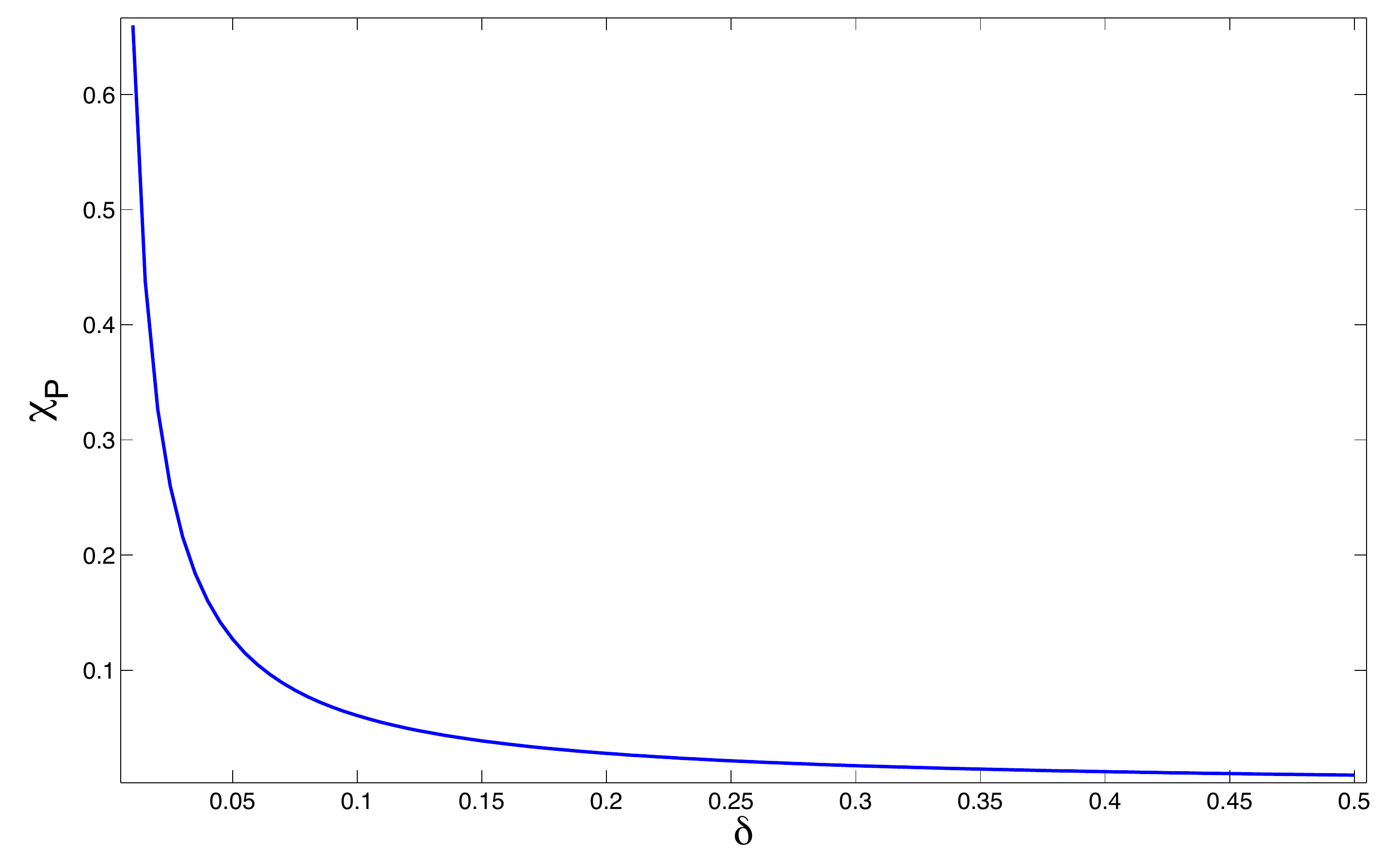}
\caption{(Color online) Ratio between the electrostatic force at second order in the corrugation and the corresponding force at zeroth order ($\chi_{\rm P} = \frac{{\mathcal F}_{P,2}}{{\mathcal F}_{P,0}}$ ), as a function of the distance $\delta = d/R$, for the sphere-plane geometry. Parameters are: $p_{\rm max} R = 10000$, $p_{\rm min} R = 100$, and $\epsilon/R = 0.001$.}
\label{fuerza2E}
\end{figure}

We have numerically evaluated the effect of the roughness on the electrostatic force for the sphere-plane geometry, for the 
sharp-cutoff model. The result is shown in Fig.\ref{fuerza2E}, where the plot shows the ratio $\chi_P$  between the force at second order 
and the one at zeroth order in the corrugation.
As expected, the effect of the roughness is more relevant at short distances. The plot shows a $1/d$ behaviour for small $d$, 
compatible with the previous analytic results for large values of $p_{\rm max}$.

\section{Results for the Casimir effect}\label{sec:casimir}
\subsection{Dirichlet boundary conditions}\label{ssec:dirichlet}
In the Dirichlet Casimir effect there is no dimensional parameter in the
problem coming from the expansion of $F$, except from $a$. 
In particular, this means that both ${\widetilde g}^{(2)}$ and ${\widetilde
h}^{(2)}$ are {\em dimensionless\/} functions of ${\mathbf
l_\shortparallel}$.

We consider here the calculation of the function $V_2$, the {\em effective
potential\/} term in the DE to the second order in the amplitude. To that
end, we need the ${\widetilde f}^{(2)}$ kernel. This object has already
been calculated \cite{Foscoetal2011}, and the result may be put as follows: 
${\widetilde f}^{(2)}({\mathbf k}_\shortparallel) =
\big[{\widetilde f}^{(2)}(k_\shortparallel)\big]\Big|_{k_0\to 0}$, where \cite{pfaT}
\begin{equation}
{\widetilde f}^{(2)}(k_\shortparallel) \;=\; -2
\int \frac{d^3p_\shortparallel}{(2\pi)^3} \;
\frac{|p_\shortparallel| \, |p_\shortparallel + k_\shortparallel|}{(1 -
e^{- 2 a |p_\shortparallel|}) (e^{2 a |p_\shortparallel + k_\shortparallel|}-1)}
\end{equation}
where we have used the notation $l_\shortparallel \equiv (l_0,l_1,l_2)$,
and $|l_\shortparallel| \equiv \sqrt{l_0^2 + {\mathbf
l_\shortparallel}^2}$, for any $3$-vector $l_\shortparallel$.

Thus, in this case, the function ${\widetilde g}^{(2)}({\mathbf
l_\shortparallel})$ is a dimensionless function independent of any
dimensionful parameter and, in practice, it may be obtained as follows:
\begin{equation}
{\widetilde g}^{(2)}({\mathbf l_\shortparallel}) \,=\, 
\big[{\widetilde f}^{(2)}({\mathbf l_\shortparallel}) \big]\Big|_{a \to 1}
\;,
\end{equation}
which may be written explicitly as follows ($x$ = $|{\mathbf
l_\shortparallel}|$):
\begin{eqnarray}
{\widetilde g}^{(2)}(x) &=&
\frac{x^3 \text{Li}_2\left(e^{-2 x}\right)}{48 \pi ^2}+\frac{x^2
\text{Li}_3\left(e^{-2 x}\right)}{24 \pi ^2}+\frac{x \text{Li}_4\left(e^{-2
x}\right)}{16 \pi ^2}+\frac{\text{Li}_5\left(e^{-2 x}\right)}{16 \pi
^2}+\frac{\pi ^2 \text{Li}_2\left(1-e^{-2 x}\right)}{240 x}\nonumber\\ 
&+& \frac{\text{Li}_6\left(e^{-2 x}\right)-\frac{\pi ^6}{945}}{32 \pi ^2
x}-\frac{x^4 \log \left(1-e^{-2 x}\right)}{120 \pi ^2}+\frac{\pi ^2
x}{240},
\end{eqnarray}
where ${\rm Li}_n(x)$ denote the Polylogarithm functions. 

\begin{figure}[!ht]
\includegraphics[width=11.0cm]{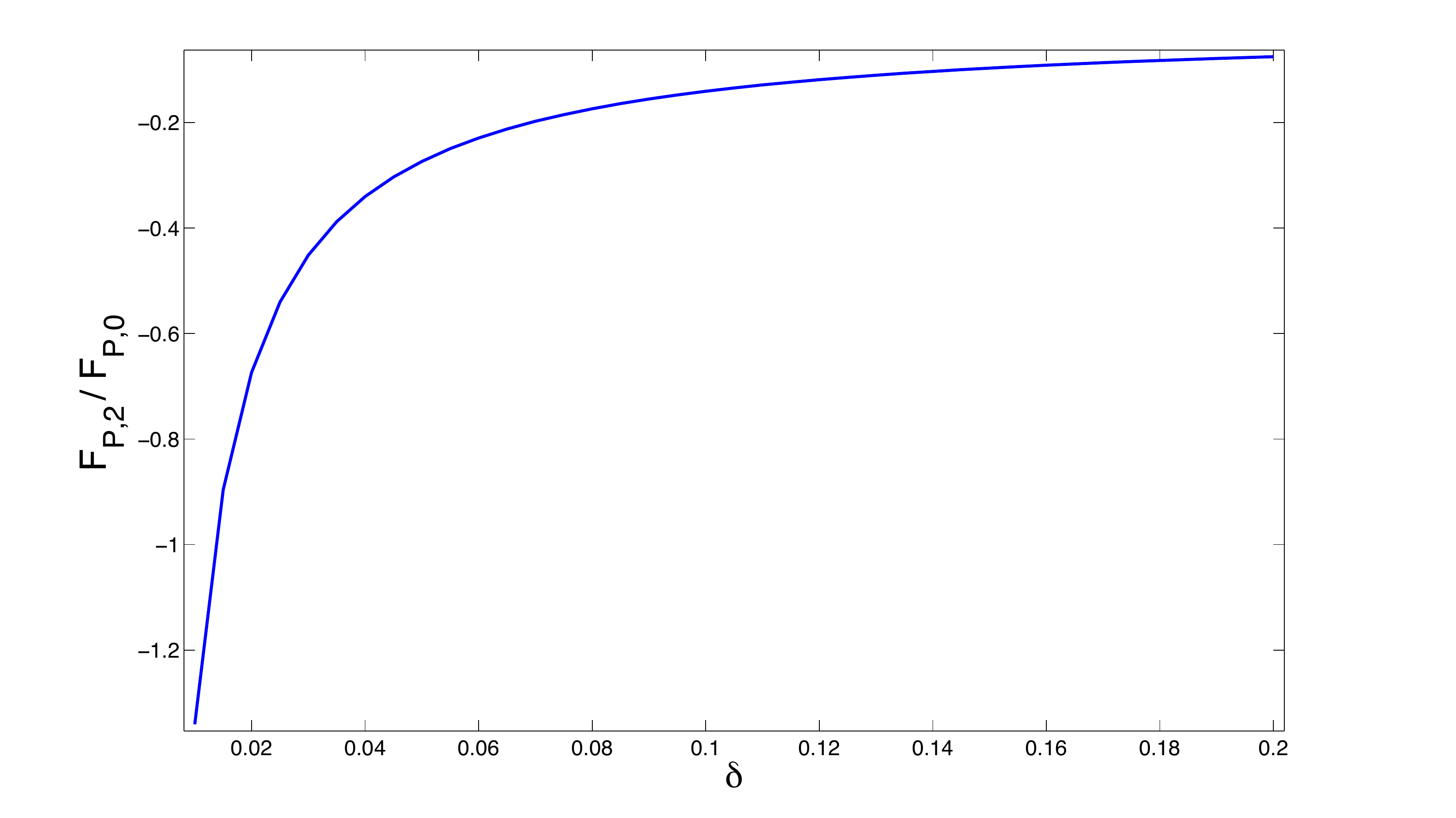}
\caption{(Color online) Ratio between the Dirichlet Casimir energy at second in the corrugation, and zeroth order as a function 
of the distance $\delta = d/R$, for the sphere-plane geometry.
 Parameters are: $p_{\rm max} R = 10000$, $p_{\rm min} R = 100$, and $\epsilon/R = 0.001$.}
\label{F2D}
\end{figure}

In Fig.\ref{F2D} we show the numerical evaluation of Eq.(\ref{F2}) for the Dirichlet Casimir energy for the sphere-plane geometry,
using the sharp cutoff model for the roughness. In this figure we plot the ratio between the second and 
zeroth order in the corrugation $F_{P,2}/F_{P,0}$ as a function of the minimal distance between the sphere and the plane.
The $1/d$ behaviour is similar to the one of the electrostatic case, and can be derived analytically for large values of $p_{\rm max}$
 taking into account the large $x$ limit of  $ {\widetilde g}^{(2)}(x)$. 

\subsection{Neumann boundary conditions}\label{ssec:neumann}

\begin{figure}[!ht]
\includegraphics[width=10.0cm]{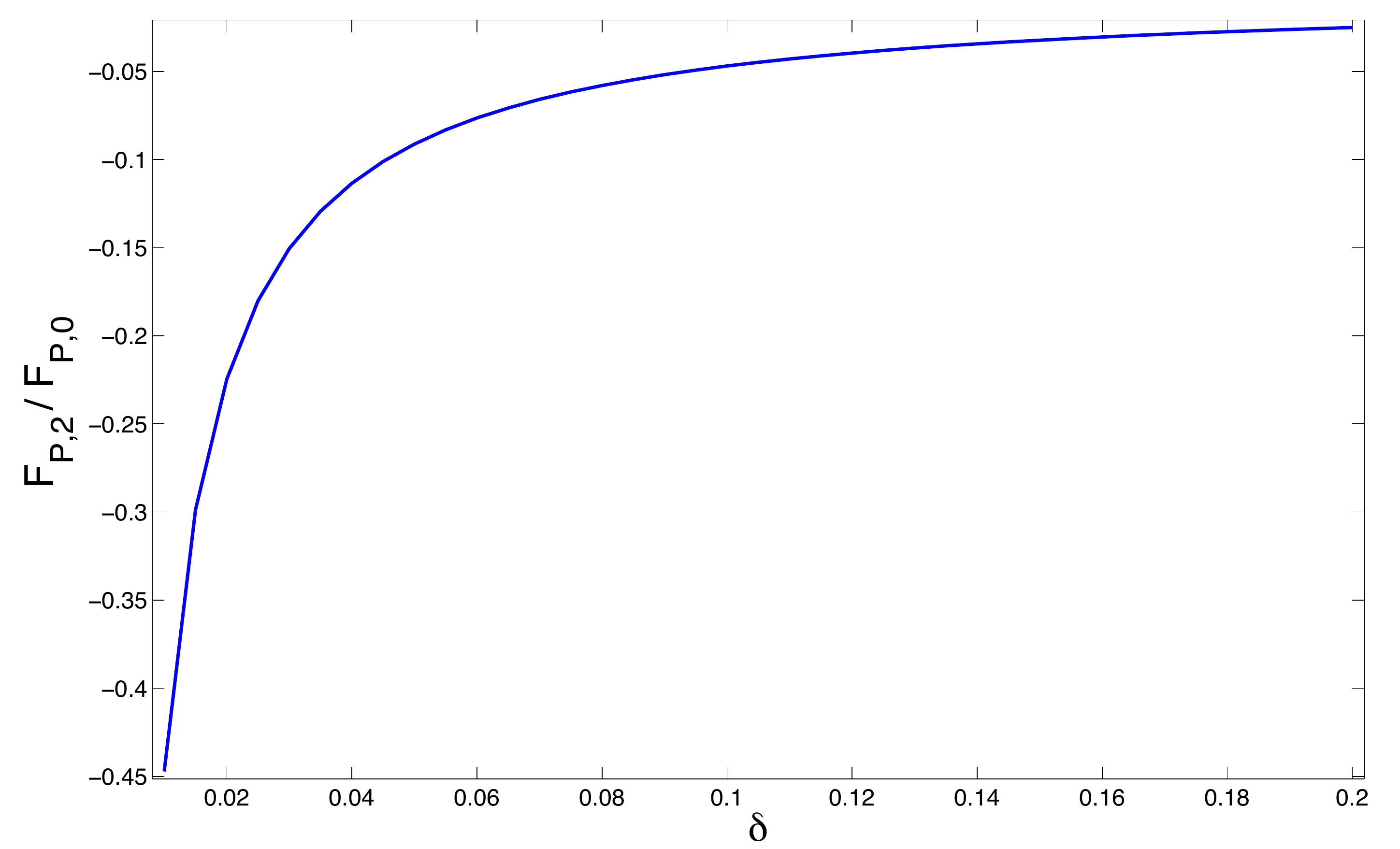}
\caption{(Color online) Ratio between the Neumann Casimir energy at second and zeroth order in the corrugation as a function 
of the distance $\delta = d/R$, for the sphere-plane geometry.
 Parameters are: $p_{\rm max} R = 10000$, $p_{\rm min} R = 100$, and $\epsilon/R = 0.001$.}
\label{Fig2N}
\end{figure} 

In the Neumann Casimir effect, the function 
${\widetilde f}^{(2)}$ is given by \cite{pfaT}:
\begin{eqnarray}
{\widetilde f}^{(2)}(k_\shortparallel)  = - 2\, \int \frac{d^3p_\shortparallel}{(2\pi)^3}
	\frac{\big[p_\shortparallel \cdot (p_\shortparallel + k_\shortparallel)\big]^2}{|p_\shortparallel| \, 
	|p_\shortparallel + k_\shortparallel|} \, 
\frac{1}{1 - e^{- 2a |p_\shortparallel|}}  \frac{1}{e^{2a |p_\shortparallel + k_\shortparallel|} - 1 } \;.
\label{F2N}
\end{eqnarray}
and therefore

\begin{eqnarray}
&& {\widetilde g}^{(2)}(x) =  -\frac{1}{24} \left(\frac{x^2}{2 \pi ^2}+1\right) x \text{Li}_2\left(e^{-2
   x}\right)+\left(\frac{x^2}{48 \pi ^2}-\frac{1}{16}\right)
   \text{Li}_3\left(e^{-2 x}\right) \nonumber\\
&+&\frac{5 x \text{Li}_4\left(e^{-2
   x}\right)}{32 \pi ^2}+\frac{7 \text{Li}_5\left(e^{-2 x}\right)}{32 \pi
   ^2} +\frac{\pi ^2 \text{Li}_2\left(1-e^{-2 x}\right)}{240 x} \nonumber\\
&+&\frac{-\pi ^2
   \text{Li}_4\left(e^{-2 x}\right)+\frac{7 \text{Li}_6\left(e^{-2
   x}\right)}{2}+\frac{\pi ^6}{135}}{32 \pi ^2 x}-\frac{x^4 \log \left(1-e^{-2
   x}\right)}{120 \pi ^2}+\frac{\pi ^2 x}{720}\, .
\end{eqnarray}
In Fig. \ref{Fig2N} we show the numerical result of evaluating Eq.(\ref{F2}) for the Neumann Casimir energy for the sphere-plane geometry,
using the sharp cutoff model for the roughness. Once again, the effect of the roughness becomes relevant at short distances.

\section{Conclusions}\label{sec:concl}
We have found general expressions for the DE of the
interaction energy between two surfaces, including the first nontrivial
correction due to corrugation, assumed to exist on top of an otherwise
smooth surface facing a plane. 

The procedure we have followed to compute the interaction energy between
surfaces is conceptually very simple. Due to the roughness, the functional
that describes the interaction energy between surfaces does not admit an
expansion in derivatives. However, after averaging  over the small scale
fluctuations, the resulting functional only depends on the shape of the
smooth surfaces (the spatial averages of the rough ones), and therefore it
makes sense to compute it using a DE. The leading order correction $F_{\rm
P,2}$ can be thought as the usual PFA applied to the effective interaction
that takes into account the roughness on parallel plates,
while the next to leading term improves that by adding corrections
depending on the derivatives of the function which defines the curved
surface.

Note that, although the roughness is assumed to be described as a random variable, its statistical properties become
inextricably mixed with the geometry of the surface, even at the first
non-trivial order in the amplitude. At this order, the two-point
correlation function is all the information that one needs to know.   
It should be clear that the general results could be applied the other way
around, namely, one could attempt to determine the characteristics of the roughness of a surface 
by performing force measurements.

We have applied the general results to the case of the interaction between
a rough sphere and a plane, both for electrostatic and Casimir
interactions. We have considered the particular case in which roughness can
be described by a simple correlation function which is constant between the
bandwidth set up by two momentum space cutoffs. Of course the results can be extended to more
realistic correlation functions, including roughness described by
self-affine fractal scaling.  

The results of this paper could also be
generalized in other directions, like for instance to the case of two
curved surfaces having roughness and finite conductivity.

\section*{Acknowledgements}
This work was supported by ANPCyT, CONICET, UBA and UNCuyo. We are grateful to C. Garc\'\i a Canal for suggesting us the study of a related problem, 
what  lead us to the research presented in this paper. We are also appreciative of his encouraging comments on our manuscript.

\end{document}